\documentclass{svmult}

\usepackage{makeidx}         
\usepackage{graphicx}        
\usepackage{multicol}        
\usepackage[bottom]{footmisc}

\makeindex             

\begin{document}

\title*{Radiative transfer problem in dusty galaxies: ray-tracing approach}
\titlerunning{Radiative transfer problem in dusty galaxies}
\author{
Dmitrij Semionov
\and Vladas Vansevi\v{c}ius
}
\institute{Institute of Physics, Savanori\c{u} 231, LT-03154 Vilnius,
Lithuania \texttt{dima@itpa.lt}}

\maketitle

\begin{abstract}
A new code for evaluation of light absorption and scattering by
interstellar dust grains is presented. The radiative transfer
problem is solved using ray-tracing algorithm in a self-consistent
and highly efficient way. The code demonstrates performance and
accuracy similar or better than that of previously published results,
achieved using Monte-Carlo methods, with accuracy better than
$\sim$ 3\% for most cases. The intended application of the code is
spectrophotometric modelling of disk galaxies, however, it can be
easily adapted to other cases that require a detailed spatial
evaluation of scattering, such as circumstellar disks and shells
containing both point and distributed light sources.
\end{abstract}

\section{Problem statement}
\label{sec:1}

The purpose for the developing radiative transfer problem solving
code, described in this article, was to model spatial and spectral
energy distribution (SED) observed in external galaxies. The nature
of this problem requires `self-consistency' of a solution -- the
resulting SED of a model must depend only on the SED of stellar
sources and assumed properties of dust without any preconditioning
on light and attenuation distribution within galaxy \cite{tvapaper}.

While galaxies in general are complex objects with three-dimensional
(3D) distribution of radiation and mater, in most cases they are
dominated by axial symmetry (2D), allowing significant simplification
of the model geometry. However, the model should account for presence
of macroscopic structure within galaxies, possibly including elements
having other symmetry, such as bars and spiral arms (2D+).

Most present day astrophysical radiative transfer codes employ
either a Monte-Carlo (MC, eg. \cite{cfmg}) or a ray-tracing (RTR, eg.
\cite{razoumovscott}, \cite{rmas}) methods. Some of implementations
of these methods were compared by \cite{bdj} for 1D and by
\cite{dullemond} for 2D cases. The RTR approach allows the
optimization of solution for a given system geometry, which was
the main reason to use it as a basis for the Galactic Fog Engine
(hereafter `GFE'), a program for self-consistent solution of
radiative transfer problem in dusty media with primarily
axisymmetric geometry. This paper concentrates on radiative transfer 
in ultraviolet-to-optical wavelength range assuming exclusively 
coherent scattering, therefore in most equations the wavelength 
dependence will be omitted.

\section{Algorithm}
\label{sec:2}

\subsection{Model description}

The GFE iteratively solves a discrete bidirectional radiative 
transfer problem, producing intensity maps of the model under 
arbitrary inclination at a given wavelength set. The foundation 
of the iterative evaluation of radiation transfer equation was 
laid out by Henyey \cite{henyey}. By solving a set of one-dimensional 
radiative transfer equations 

\begin{equation}
{dI \over ds} = - \kappa I + j + \kappa {\omega \over 4 \pi}
\int I \Phi d \Omega,
\end{equation}

\noindent where $\Phi$ and $\omega$ denote the scattering phase 
function and albedo and $\kappa$ and $j$ are absorption and 
emissivity coefficients of the medium, the initial system SED 
is separated into: escaped energy, that reaches an external 
observer; energy, absorbed by grains, that is eventually emitted 
as thermal radiation; and scattered energy distribution. The 
solution is then repeated, substituting scattered energy as 
initial distribution for the next iteration, accumulating 
resulting escaped and absorbed energy, until certain convergence 
criteria are met, either a fixed number of iterations, or 
remaining scattered energy being below specified threshold. 
After convergence is reached the dust temperature is calculated 
from absorbed energy distribution. If it is necessary to account 
for self-scattering of thermal radiation by dust grains, the 
resulting emission SED can be input into scattering evaluation 
loop and the process repeated until the final radiative energy 
distribution is obtained, and then used to produce SED as seen 
by an external observer. 

\begin{figure}
\centering
\includegraphics[height=7.5cm]{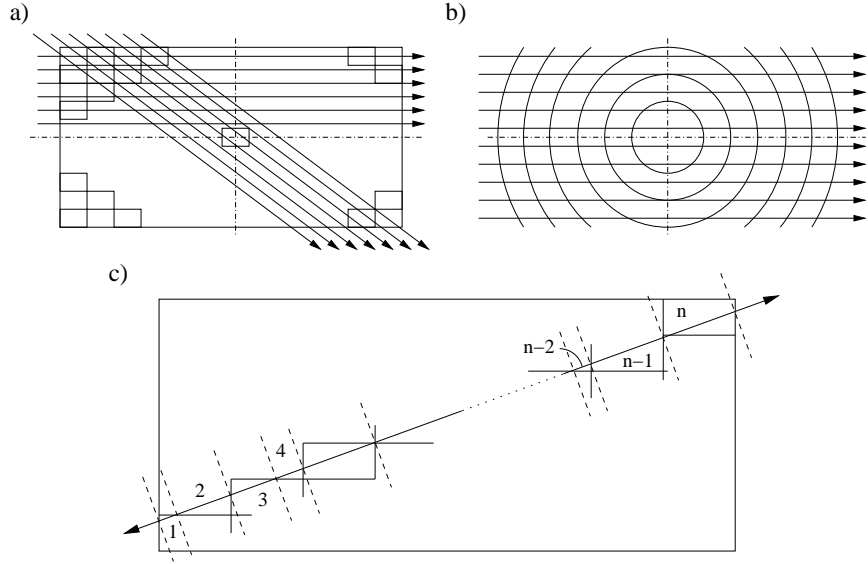}
\caption{The model geometry. Panel a) shows the diametral, panel
b) - central plane cross-section of the model. The distribution
and density of ray-tracing paths (shown as arrows) are computed to
produce even and sufficient sampling of the model volume. Panel c)
illustrates the discrete radiative transfer in cross-section
parallel to the model $Z$ axis. Limits of plane-parallel layers
for one-directional treatment are shown as dotted lines while
boundaries of the individual rings are outlined with solid lines.}
\label{fig:1}
\end{figure}

Calculations are performed within a cylinder with a radius $r_m$
and height above midplane $z_m$, which is subdivided into a set of
layers of concentric, internally homogeneous rings (`bins') of
arbitrary radial and vertical thickness. Since the linear extent
of each individual light source (star) is negligible compared to
the size of system, it is possible to solve the radiative transfer
problem considering every volume element of the model having both
attenuating (light scattering and absorption by interstellar dust)
and emitting (light emission by the stars and thermal radiation of
the dust particles) properties per unit volume, defining for each 
ring denoted by indexes $r$ and $z$ its total absorption 
$k' = \kappa(r,z)$, and emissivity $j' = j(r,z,\alpha,\delta)$, 
combined from internal light sources and energy scattered within 
its volume, with angles $\alpha$ and $\delta$ defining the direction 
of radiation propagation.

\subsection{Radiative transfer equation}

GFE uses static ray-casting geometry, determining the set of rays
that ensures a required degree of sampling of the model volume
(fig.~\ref{fig:1}a and \ref{fig:1}b). If the viewing solid angle, 
containing each direction, can be held small, the radiative transfer 
along these rays can be solved as in plane-parallel homogeneous 
layer case for a series of intervals traversing rings until crossing 
the outer boundary of the model (fig.~\ref{fig:1}c). In a form 
suitable for computer implementation an incident intensity on a 
given point for a light path separated into $n$ intervals of length 
$l_i$, numbered outwards from that point, is written as

\begin{equation}
I_{\rm inc}=\sum_{i=1}^n \left(\prod_{j=1}^{i-1}e^{-k'_j l_j}\right)
{j'_i \over k'_i} \left(1-e^{-k'_i l_i}\right).
\end{equation}

Similarly, the intensity of radiation scattered with albedo $\omega$
from a given direction $(\alpha : \delta)$ into all other directions
within a certain interval denoted by index ``1'' is

\begin{equation}
I_{1,\alpha,\delta}=
    \omega j'_1 l_1 - \omega \left( 1 - e^{-k'_1 l_1} \right)
    \left[
    {j'_1 \over k'_1} - \sum_{i=2}^n \left( \prod_{j=2}^{i-1} e^{-k'_j l_j} \right)
    {j'_i \over k'_i} \left(1-e^{-k'_i l_i}\right)
    \right].
\end{equation}

When considering azimuthally inhomogeneous model configuration (3D
case), each ring is subdivided into required number of azimuthal
segments. The number and directions of rays cast through the
system have to be modified accordingly to include new sets of rays
in azimuthal direction, however, the ray-tracing part of the
algorithm is unchanged. The computational time scales as $N_{\rm
bin}^{3/2} \times \log N_{\rm bin}$ for 2D and $N_{\rm bin}^{4/3}
\times \log N_{\rm bin}$ for 3D cases.

\subsection{Scattering phase function}

Since angular distribution of radiation at each point in the model
is non-isotropic, it must be described using a numerical phase
function (matrix), providing the radiation intensity towards a set
of predefined reference directions (`RDs') described by angular
coordinates $(\alpha_0 : \delta_0)$. There exists a number of ways
to distribute RDs on a sphere, however, those methods that produce
a set of RDs arranged in iso-latitude rows are the most efficient
in this particular model geometry, allowing both efficient storage
and retrieval of scattered intensity and fast rotation of the phase
matrix around model $Z$-axis. The memory requirements and the
overall algorithm's performance have also to be taken into
consideration.

\begin{figure}
\centering
\includegraphics[height=9cm]{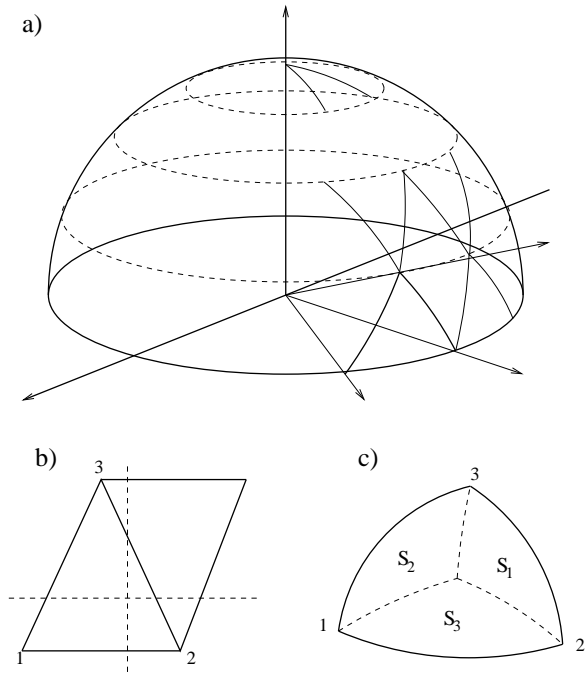}
\caption{The reference point structure used for interpolation of
the scattering phase function. Panel a) shows the trivial
iso-latitude triangulation for one hemisphere with reference
directions arranged symmetrically against diametral and horizontal
planes. Panels b) and c) represent two implemented interpolation
schemes, `flat' and `spherical'. In case of spherical
interpolation, input from each triangle vertex is weighted by the
area defined by shortest distances from the given direction to the
vertices ($S_1$ for 1-st vertex and so on).} \label{fig:2}
\end{figure}

In this work the following methods of RD distribution were compared:
HEALPix\footnote{\tt http://www.eso.org/science/healpix/} \cite{healpixpaper},
HTM\footnote{\tt http://www.sdss.jhu.edu/htm/} \cite{htmpaper},
a trivial iso-latitude triangulation (hereafter `TT'), fig.~\ref{fig:2}a)
and a square matrix with elements (`texels') corresponding to evenly
spaced $(\alpha_0 : \delta_0)$ coordinates (hereafter `Texel'). For
triangulation schemes and HEALPix the radiation intensity towards a
given point was interpolated between 3 nearest RDs using either `flat'
(fig.~\ref{fig:2}b) or `spherical' (fig.~\ref{fig:2}c) weights.

\section{Computational precision}
\label{sec:3}

\subsection{Scattering phase function interpolation}

To compare used sphere subdivision and interpolation algorithms a
following test model (hereafter a `standard model') was employed:
a cylinder with height to radius ratio $z_m / r_m = 0.2$, divided
into $N_{\rm bin} = 441$ ($21 \times 21$) equally spaced rings,
filled with radiating and absorbing particles whose density
follows a double exponential law

\begin{equation}
\rho(r,z) = \rho_0 e^{-r/r_0} e^{-z/z_0}
\end{equation}

\noindent with $r_0 = 0.2 r_m$ and $z_0 = 0.2 z_m$ and the model 
central optical depth perpendicularly to the central plane 
$\tau_{\rm ct} = 25$. The Henyey \& Greenstein \cite{hgglam} 
scattering phase function

\begin{equation}
\Phi(\theta)={{1-g^2}\over{\left(1+g^2-2 g \cos \theta\right)^{3/2}}}
\end{equation}

\noindent was used with asymmetry parameter $g = 0.75$.

\begin{figure}
\centering
\includegraphics[height=16cm]{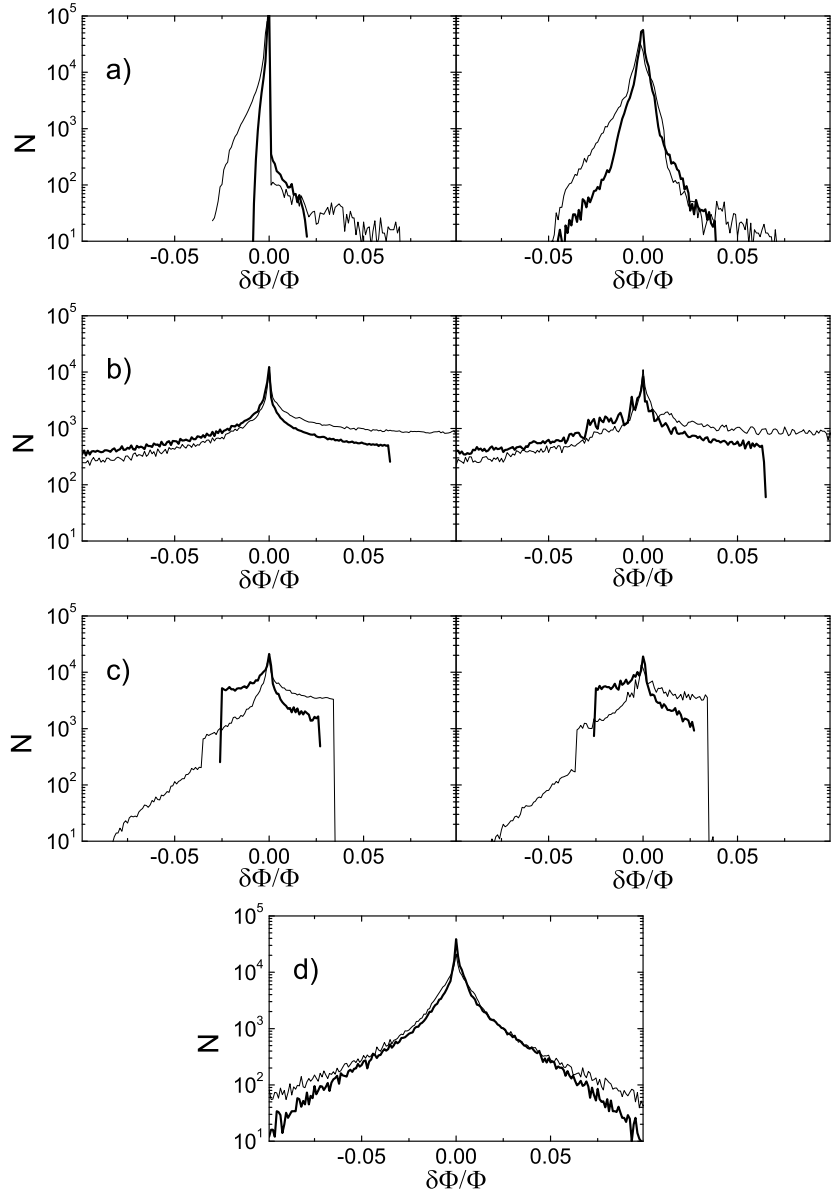}
\caption{The distribution of relative numerical phase matrix
deviation from its analytical form for different sphere
subdivision algorithms. Panels a) -- d) correspond to TT, HTM,
HEALPix and Texel methods. For the first three methods the results
obtained using both `flat' (left panels) and `spherical' (right
panels) weighted interpolation are presented. Thin line shows the
results obtained for approximately 3100, thick line -- for
approximately 12000 reference directions.} \label{fig:3}
\end{figure}

The primary quality criteria of a given algorithm is its ability
to represent the angular intensity distribution of anisotropic
scattering. If the phase function representation is exact, the
distribution of values $(\Phi'(\theta) - \Phi(\theta)) /
\Phi(\theta)$ (where $\Phi'(\theta)$ is a resulting numerical
phase matrix) would be a $\delta$-function. However, since employed
methods introduce different types of numerical errors, the actual
distribution form depends strongly on the phase function sampling
and the interpolation algorithm. An examples of resulting error
distributions (as relative numerical phase matrix deviation from
its analytical form) for the algorithms tested are shown on
fig.~\ref{fig:3}.

As can be seen, methods providing uniform sphere coverage produce more
preferable error distributions. With the increasing number of RDs the
representation of the scattering phase function improves, reducing
maximal possible deviation from the true value, particularly for TT
(fig.~\ref{fig:3}a) and HEALPix (fig.~\ref{fig:3}c) methods, with
TT algorithm showing slightly better error distribution form. `Spherical'
interpolation scheme (fig.~\ref{fig:3}, right column in panels a -- c)
produces more symmetrical error distributions, while `flat' interpolation
(left column) in some cases introduces additional numerical errors.
When compared with other methods, attempt to reproduce scattering phase
function using simple matrix with no interpolation between its elements
(Texel scheme, fig.~\ref{fig:3}d) produces results of average quality,
its error distribution quickly reaching a `saturated' form with increasing
number of RDs. The described error distribution is somewhat dependent on
the orientation of the scattering phase function relative to the set of
RDs, this dependence being minimal for the methods with identical size
of interpolation elements (HEALPix). All methods that use interpolation
display similar performance for a given number of RDs (table 1).

\begin{table}
\centering \caption{The normalized computing time for models using
different sphere subdivision and scattering phase function interpolation
algorithms.}
\label{tab:1}
\def\H{\makebox[5.2mm]{}}
\begin{tabular}{lcccc}
\hline\noalign{\smallskip}
Interpolation \H & \H HEALPix \H & \H HTM \H & \H TT \H & \H Texel  \\
method  &  &  &  &  \\
\noalign{\smallskip}\hline\noalign{\smallskip}
`flat'      & 2.5 & 2.8 & 2.5 & 1.0  \\
`spherical' & 8   & 10  & 8   & --   \\
\noalign{\smallskip}\hline
\end{tabular}
\end{table}

\subsection{Volume sampling and subdivision}

The problem encountered applying numerical methods is error
accumulation. In case of iterative ray-tracing it arises from
sampling and interpolation errors. The source of sampling errors
is incomplete/inadequate spatial sampling of the system while 
interpolation errors are related to the scattering phase function 
approximation, described in the previous section. Both error 
types independently affect every ray traced through the system, 
thus the accumulated error increases with the increasing number 
of bins and rays. This makes oversampling undesirable not only 
due to increasing computational time, but also for a reason 
of minimizing numerical errors.

As a measure of method's quality a defect in energy balance
$E_{\rm err}$ as a percentage of total energy radiated within system
$E_{\rm tot}$

\begin{equation}
E_{\rm err} = {E_{\rm tot} - E_{\rm abs} - E_{\rm sca} - E_{\rm esc} \over E_{\rm tot}}
\end{equation}

\noindent is used. Here $E_{\rm esc}$, $E_{\rm abs}$ and $E_{\rm
sca}$ are the parts of a total radiated energy that escaped the
system, was absorbed and remained to be scattered within the
system, respectively.

\begin{figure}
\centering
\includegraphics[height=5.5cm]{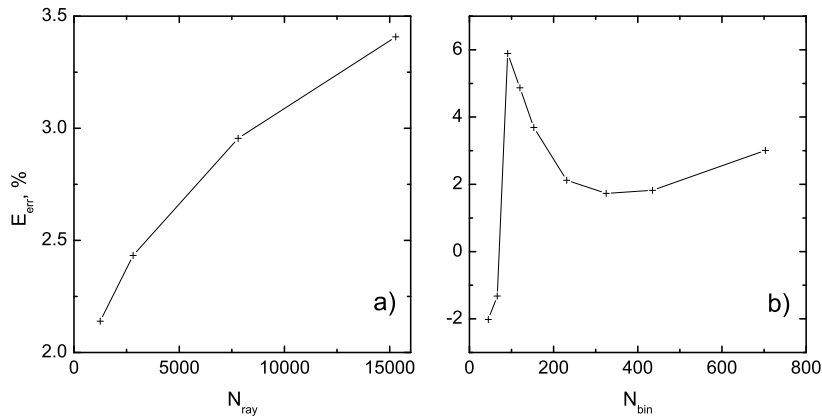}
\caption{The influence of subdivision and sampling of the model on
the energy balance. Panel a) displays the energy defect for a given
number of rays cast per model bin; panel b) shows the same
quantity for a models consisting of different number of bins.}
\label{fig:4}
\end{figure}

To determine the sampling and gridding influence on the model precision
the following two tests were performed. Firstly, the radiation field in
the standard model using TT algorithm with scattering phase function
represented by a set of 182 RDs was computed with different number of
rays $N_{\rm ray}$, cast through each ring. The results, presented on
fig.~\ref{fig:4}a, show a significant error accumulation effect. Then,
keeping a number of rays per ring constant the number of rings $N_{\rm bin}$
in model was changed (fig.~\ref{fig:4}b). As can be seen, improving the 
sampling of a model decreases the approximation errors to a certain 
minimum, limited by internal errors of a chosen scattering phase 
function interpolation method. This geometric configuration can be 
considered optimal, since with further increase in a number of bins 
and rays the quality of the solution begins to deteriorate due to 
error accumulation.

\subsection{Dust optical properties}

Other important aspect of a numerical radiative transfer solution
is its sensitivity to variations in scattering parameters: albedo
$\omega$ and asymmetry parameter $g$. Model precision and
stability for different $\omega$ and $g$ values place a constraint
on the wavelength range where a given method can be applied. The
influence of scattering parameters on energy defect was analyzed 
using the standard model with $N_{\rm bin} = 441$ and $\tau_{\rm ct} = 10$.

\begin{figure}
\centering
\includegraphics[height=5.5cm]{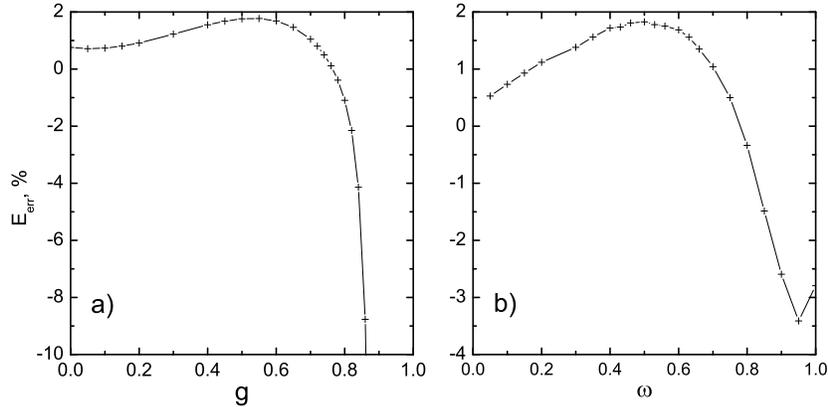}
\caption{The dependence of the energy losses within model on the
grain scattering parameters: scattering phase function asymmetry $g$
(panel a), with albedo assumed to be $\omega = 0.5$ for all $g$ values),
and albedo $\omega$ for $g = 0.6$ (panel b).}
\label{fig:5}
\end{figure}

The dependence of overall model precision on scattering asymmetry
parameter $g$ is shown on fig.~\ref{fig:5}a. The total energy defect
after 9 iterations show some variation with the increasing $g$ up to
the limit imposed by the angular scattering phase function gridding
(182 RDs) used in the calculations, after which the energy losses in
the scattering phase matrix render results invalid.

The effects of the grain albedo on the model accuracy and stability
were investigated using similar method. All computations were performed
for 7 iterations assuming $g = 0.6$. The results are shown in
fig.~\ref{fig:5}b. Within the range of $\omega$ values, applicable
to astrophysical dust grains, those errors stay in acceptable limits,
and do not influence the stability of the solution.


\section{Summary}
\label{sec:4}

The code described in this paper has undergone an extensive
testing and shows the flexibility and performance satisfying the
requirements for the models of the global radiation transfer in
dusty galaxies \cite{mcrtrpaper}. It has been successfully
applied to model both integral and position dependent SEDs of
several galaxies, some of the first results presented in
\cite{andropaper}.

The main limiting factor affecting the applicability of the
described code is the scattering asymmetry parameter $g_\lambda$.
In order to correctly treat the scattering with $g_\lambda$
approaching 1, the number of required reference directions
rises sharply, affecting both the performance and the
precision of the method. For a disk galaxy model such as one 
decribed in this paper a satisfactory convergence is obtained 
for $g_\lambda$ in [ -0.8 ; 0.8 ] range which includes the 
optical properties of typical astrophysical grains scattering 
photons from microwave up to extreme UV wavelengths.

Other model properties, such as optical depth $\tau_\lambda$ and 
the relative amount of scattered radiation (dependent on albedo 
$\omega_\lambda$) seem to have relatively little effect on the 
quality of the solution. However, models with large optical 
depth (of order of a few 100's), particularly having a steep 
matter distribution gradients, require a significant amount of 
computing time and storage. 

The application of this code is not restricted to the systems
with dispersed sources and absorbers, the algorithm being easily
extended to include treatment of interaction between radiation
field and surfaces of macroscopic objects.


\subparagraph{Acknowledgements} This work was supported by a 
Grant of the Lithuanian State Science and Studies Foundation. 
Some of the results in this paper have been derived using 
HEALPix \cite{healpixpaper} package.

%
%
%

\def\araa{Ann.\ Rev.\ Astron.\ Astrophys.}%
\def\apj{Astrophys.~J.}%
\def\apjs{Astrophys.~J.\ Suppl.\ Ser.}%
\def\aap{Astron.\ Astrophys.}%
\def\aaps{Astron.\ Astrophys.\ Suppl.}%
\def\aj{Astron.~J.}%
\def\mnras{Mon.\ Not.\ Roy.\ Astron.\ Soc.}%
\def\jqsrt{J.~Quant.\ Spectrosc.\ Radiat.\ Transfer}%



\end{document}